# Comparison of the dense molecular gas in the LINER galaxy NGC 6764 and in the Wolf-Rayet galaxy NGC 5430

T. Contini and E. Davoust

*URA 285, Observatoire Midi-Pyrénées, 14 avenue E. Belin, F-31400 Toulouse, France*

H. Wozniak

*Observatoire de Marseille, 2 place Le Verrier, F-13248 Marseille Cedex 4, France*

S. Considère

*Observatoire de Besançon, B.P. 1615, F-25010 Besançon, France*

**Abstract.** We compare the dense molecular gas content in two barred spiral galaxies, NGC 6764 (classified LINER) and NGC 5430 (a Wolf-Rayet galaxy). We find a significant difference in the proportion of dense molecular gas between the two galaxies. $CS(3 \rightarrow 2)$ is detected in NGC 6764 but not in NGC 5430, even though the intensities of $^{12}CO(2 \rightarrow 1)$ and $HCN(1 \rightarrow 0)$ are higher in the latter galaxy. The non detection in NGC 5430 indicates that the CS abundance in that galaxy is unusually low, or that HCN is subthermally excited. To complement these observations, we discuss the ionization source in the nucleus of the LINER galaxy NGC 6764.

## 1. Introduction

It is now well established (Noguchi 1988, Friedli & Benz 1993, Wada & Habe 1995) that bars play an important role in the process of nuclear activity (starburst or AGN), by efficiently funnelling molecular clouds toward the center of galaxies. This process leads to the accumulation of a large concentration of molecular gas in the nucleus of galaxies, where most starbursts are observed to occur.

The CO molecule has been very extensively used as a tracer of molecular clouds, by the observation of several transitions of the molecule in the millimeter domain. However, it has recently been emphasized that CO only traces low-density gas ($n(H_2) \leq 10^3$ cm$^{-3}$), and several searches for dense ($n(H_2) \geq 10^3$ cm$^{-3}$) molecular clouds in normal and active galaxies, via detection of HCN, CS, HCO$^+$, and other molecules, have been initiated (e.g. Mauersberger et al. 1989, Nguyen-Q-Rieu et al. 1992, Helfer & Blitz 1993).

Among other results, these surveys have shown that bulges of normal as well as starburst galaxies contain large quantities of dense gas, that HCN is brighter than CS by a factor 2.4 in normal galaxies (Helfer & Blitz 1993), and that the mass of $H_2$ is reasonably estimated by the standard $CO - H_2$ relation, even in starburst galaxies (Sage et al. 1990).

In order to better understand the relationship between LINER activity and the process of intense star formation in barred spiral galaxies, we compare the abundance of dense molecular clouds (via the detection of the HCN and CS molecules) in the LINER galaxy NGC 6764, and in the Wolf-Rayet galaxy NGC 5430. We choose these two galaxies because they have nearly the same fundamental properties, such as morphological type, size, luminosity or inclination (see Table 1).

Table 1.   Fundamental properties of the barred galaxies NGC 6764 and NGC 5430

| Properties | NGC 6764 | NGC 5430 |
|---|---|---|
| Right ascension (1950) | $19^h\ 07^m\ 01.5^s$ | $13^h\ 59^m\ 08.5^s$ |
| Declination (1950) | $50^\circ\ 51'\ 03''$ | $59^\circ\ 34'\ 10''$ |
| Morphological type | SBbc | SBb |
| Apparent blue magnitude | 12.68 | 12.84 |
| Absolute magnitude | -20.9 | -20.7 |
| Distance ($H_0 = 75\ kms^{-1}Mpc^{-1}$) | 35.6 Mpc | 42.7 Mpc |
| Inclination | $58.1^\circ$ | $51.7^\circ$ |
| Velocity (HI) | $2416\ kms^{-1}$ | $2965\ kms^{-1}$ |
| $W_{50}$ (HI) | $276\ kms^{-1}$ | $313\ kms^{-1}$ |
| $\log(M_{HI})$ | $9.65\ M_\odot$ | $9.58\ M_\odot$ |
| $\log(L_B)$ | $10.52\ L_\odot$ | $10.46\ L_\odot$ |
| $\log(L_{fir})$ | $10.23\ L_\odot$ | $10.63\ L_\odot$ |
| $\log(M_{HI}/L_B)$ | -0.87 | -0.88 |
| $\log(L_{fir}/L_B)$ | -0.29 | 0.17 |

## 2.   Observations

The millimeter observations were obtained at the IRAM 30m Radiotelescope on Pico Veleta on November 27 and 28, 1994. We observed in lower side band at three frequencies simultaneously. The configuration of the observations is summarized in Table 2. The weather was remarkably good during the run with mean opacities equal to 0.17 at 1.3mm, 0.07 at 2mm and 0.08 at 3mm.

Table 2.   Configuration of millimeter observations

| Transition | Frequency | Receiver | Beam size $\theta_b$ |
|---|---|---|---|
| HCN(1→0) | 88.632 GHz | 3mm SIS | $27''$ |
| CS(3→2) | 146.969 GHz | 2mm SIS | $16''$ |
| $^{12}CO(2→1)$ | 230.538 GHz | 1.3mm G1 SIS | $11''$ |

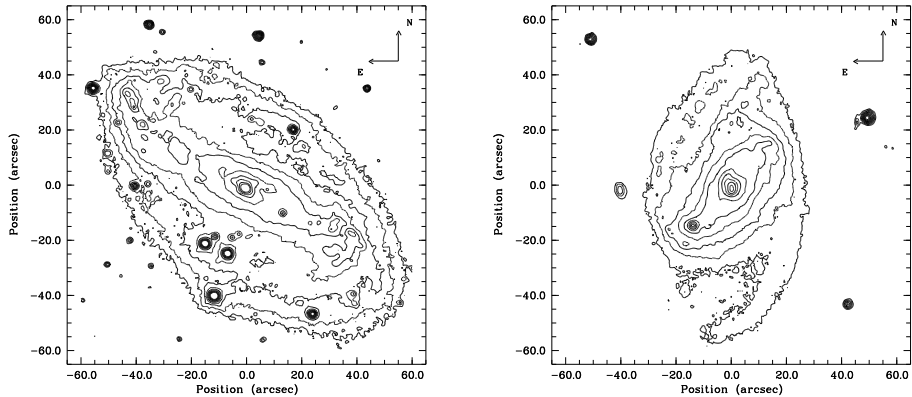

Figure 1. Isophotal contour maps in the V band of the LINER galaxy NGC 6764 (left) and the Wolf-Rayet galaxy NGC 5430 (right). In NGC 5430, the region of Wolf-Rayet stars is located at the South-Eastern end of the bar

CCD images were obtained during two runs at the 2-meter telescope of Observatoire du Pic-du-Midi, with a 1000×1000 Thomson CCD (pixel size of 0.2″ on the sky). Images of NGC 6764 were obtained in August 1992, with exposure times 20 min in V and 40 min in B. Images of NGC 5430 were obtained in January 1992, with exposure times 30 min in V and 20 min in R.

## 3. Results

### 3.1. Molecular line profiles

The molecular line profiles are presented in Figure 2. HCN(1→0) was detected in the centers of NGC 6764 and NGC 5430, but not in the Wolf-Rayet region of NGC 5430. CS(3→2) was only (marginally) detected in the center of NGC 6764. The corresponding velocities and fluxes are summarized in Table 3. Since NGC 6764 and NGC 5430 are fairly distant compared to most galaxies where HCN has been detected, the HCN fluxes given in this paper are among the lowest ever reported.

The central $^{12}CO(2→1)$ profile of NGC 5430 has previously been observed by Krügel et al. (1990), with the same instrument; the profile and total flux are in good agreement with ours. They also observed the $^{12}CO(1→0)$ and $^{13}CO(1→0)$ lines in the center of this galaxy. The central $^{12}CO(2→1)$ profile of NGC 6764 has been published by Eckart et al. (1991), also using the IRAM 30m antenna; they find a rather high total flux of 35 Kkms$^{-1}$ compared to ours. They also mapped the two transitions of CO in the galaxy.

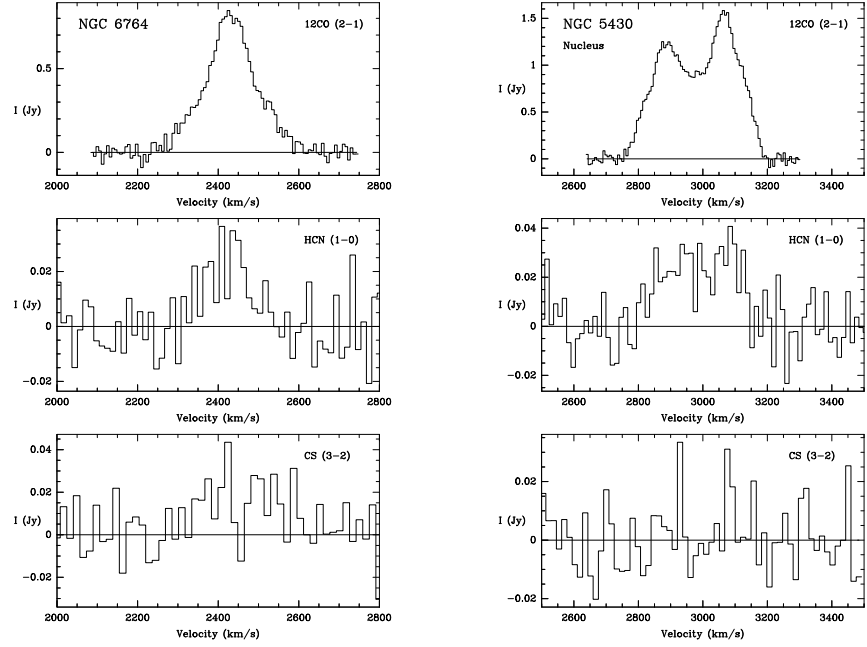

Figure 2.    Molecular line profiles (in Jy) at the centers of NGC 6764 (left) and NGC 5430 (right). Top: profile of $^{12}$CO(2→1), middle: profile of HCN(1→0), bottom: profile of CS(3→2)

Table 3.    Velocities, line widths and fluxes of the molecular lines

| Galaxy | $^{12}$CO(2→1) | | | HCN(1→0) | | | $\frac{^{12}CO(1\to0)}{HCN(1\to0)}$ |
|---|---|---|---|---|---|---|---|
| | V kms$^{-1}$ | FWHM kms$^{-1}$ | Flux Jykms$^{-1}$ | V kms$^{-1}$ | FWHM kms$^{-1}$ | Flux Jykms$^{-1}$ | |
| **NGC 6764** | 2431±1 | 148±2 | 122±2 | 2422±11 | 139±25 | 4±1 | 31 |
| **NGC 5430** Nucleus | 2981±1 | 389±3 | 389±2.5 | 2985±17 | 248±29 | 7±1 | 21 |
| W-R region | 3074±3 | 97±5 | 39±2 | | | | |

## 3.2. Proportion of dense molecular clouds

The purpose of this paper is to determine if there is a significant difference between the proportion of dense molecular gas in the LINER galaxy NGC 6764 and in the Wolf-Rayet galaxy NGC 5430. The results are summarized as follow:

- we found that the HCN intensities of both galaxies (see Table 3) are very close to the mean value $(^{12}CO(1\rightarrow0)/HCN(1\rightarrow0)=25)$ predicted by the HCN-CO relation of Nguyen-Q-Rieu et al. (1992).

- the marginal detection of CS in NGC 6764 is consistent with the mean value of intensity ratios found in the literature (Sage et al. 1990, Helfer & Blitz 1993).

- the non detection of CS in NGC 5430 indicates that the CS abundance is unusually low or that HCN is subthermally excited in that galaxy.

## 4. Origin of LINER activity in NGC 6764

The nature of the nuclear activity of NGC 6764 is not well established. It was originally classified as a Seyfert 2 galaxy by Rubin et al. (1975) and Koski (1978). Later, Heckman (1980) included NGC 6764 in his list of LINERs, on account of its large [OI]$\lambda$6300/[OIII]$\lambda$5007 ratio ($\sim 0.3$).

On the other hand, Osterbrock & Cohen (1982) noted the presence of emission features from Wolf-Rayet stars in the nucleus of NGC 6764. The same emission-lines from Wolf-Rayet stars (principally the broad HeII$\lambda$4686 line) were detected by Keel (1987), not in the nucleus, but at the end of the bar of NGC 5430. The ages of the starbursts in NGC 6764 and NGC 5430 are given in Table 4. They were estimated using spectrophotometric data from the literature (H$\beta$ equivalent width from Osterbrock & Cohen 1982, and Mazzarella & Boroson 1993 ; oxygen abundance from Kunth & Joubert 1985, and Keel 1987), and models of starburst computed by Cerviño & Mass-Hesse (1994 ; CMH94) and Leitherer & Heckman (1995 ; LH95).

The detection of Wolf-Rayet stars in the nucleus of NGC 6764 and at the end of the bar of NGC 5430 indicates that the starbursts are very young ($\leq 6$ Myr) and that many massive stars were born during the bursts. Using spatially-resolved optical spectra to determine emission-line ratios, Boer & Schulz (1990) made a distinction between the ionization source in the nucleus and in the circumnuclear region of NGC 6764. They found that the emission-line ratios [NII]$\lambda$6583/H$\alpha$ and [SII]$\lambda\lambda$6713+6731/H$\alpha$ increase from the center to the circumnuclear region whereas the emission-line ratio [OIII]$\lambda$5007 remains nearly constant. Using these ratios and the diagnostic diagrams of Veilleux & Osterbrock (1987), we found a starburst-like activity in the nucleus whereas the LINER activity originates from the circumnuclear region.

We can thus interpret the LINER activity in NGC 6764 in terms of two ionization sources: *photoionization by young and massive stars in the nucleus* (corroborated by the presence of Wolf-Rayet stars) and *excitation by shocks from supernovae in the circumnuclear region*.

Table 4.    Age of the starbursts in NGC 6764 and NGC 5430 estimated from spectrophotometric data and starburst models

| Galaxy | W(Hβ) (Å) | (O/H) (⊙) | Starburst age (Myr) LH95 | CMH94 |
|---|---|---|---|---|
| **NGC 6764** | 23 | 0.87 | 6.4 | 5.0 |
| **NGC 5430** | | | | |
| Nucleus | 5 | 0.90 | 8.0 | 9.0 |
| W-R region | 44 | 0.90 | 3.0 | 4.0 |

**Acknowledgments.**    Data from the literature were obtained with the Lyon-Meudon Extragalactic Database (LEDA) supplied by the LEDA team at the CRAL-Observatoire de Lyon (France). We thank Bertrand Lefloch for assistance at the IRAM radiotelescope.